\documentclass[10pt,conference]{IEEEtran}
\IEEEoverridecommandlockouts
\usepackage{algorithm}  
\usepackage{amsmath,amsfonts} 

\usepackage[table,xcdraw]{xcolor}
\usepackage{graphicx,amsmath,amssymb,amsfonts,cite}
\usepackage{booktabs,makecell} 
\usepackage{diagbox,tabu,enumerate,fancybox}
\usepackage{booktabs}
\usepackage{amsmath,amssymb,amsfonts}
\usepackage{algorithmic}
\usepackage{graphicx}
\usepackage{subcaption}
\usepackage{textcomp}
\usepackage{xcolor}
\usepackage{amsthm}
\usepackage{amsmath}

\def\BibTeX{{\rm B\kern-.05em{\sc i\kern-.025em b}\kern-.08em
		T\kern-.1667em\lower.7ex\hbox{E}\kern-.125emX}}
\begin{document}
	
	\title{Joint Phase Noise and Off-Grid Channel Estimation
		for AFDM Systems via Sparse Bayesian Learning}  
	\author{
		\IEEEauthorblockN{You Xu$^{1}$, Huaijin Zhang$^{1}$, Lixia Xiao$^{1}$, Guanghua Liu$^{1}$, and Zilong Liu$^{2}$}	
		\IEEEauthorblockA{$^{1}$School of Cyber Science and Engineering, Huazhong University of Science and Technology, Wuhan 430074, China}
		\IEEEauthorblockA{$^{2}$School of Computer Science and Electronics Engineering, University of Essex, UK}
		\IEEEauthorblockA{Email: youxu@ieee.org, \{huaijinzhang, lixiaxiao\}@hust.edu.cn, guanghualiu@ieee.org, zilong.liu@essex.ac.uk}
		\vspace{-2em}
	}
	\maketitle 
	\begin{abstract}
		In practical affine frequency division multiplexing (AFDM) systems, the intricate coupling of oscillator phase noise (PN) and off-grid fractional shifts traps conventional estimators in a severe high-SNR error floor. To address these challenges, we propose a joint PN and channel estimation method based on sparse Bayesian learning (JPNCE-SBL). Specifically, a reduced-rank subspace projection is first introduced to capture the dominant eigen-energy of the Wiener PN process. Concurrently, a dynamic grid evolution strategy is designed to iteratively eliminate off-grid errors without requiring computationally prohibitive global grid densification.  Both components are integrated into a unified Expectation-Maximization (EM) framework, where the channel and PN estimates are jointly updated at each iteration to prevent error propagation. Simulation results demonstrate that JPNCE-SBL significantly outperforms existing benchmarks in both NMSE and BER, closely approaching the perfect channel state information  case under practical PN conditions.
	\end{abstract}
	
	\begin{IEEEkeywords}
		Affine frequency division multiplexing, doubly-dispersive channels, phase noise, sparse Bayesian learning, grid evolution, joint estimation.
	\end{IEEEkeywords}  
	\section{Introduction} 
	Emerging 6G applications demand ultra-reliable connectivity in high-mobility scenarios \cite{saad2020vision, wang2023road}. However, extreme Doppler spreads in doubly-dispersive channels induce severe inter-carrier interference (ICI) in conventional orthogonal frequency division multiplexing (OFDM) systems, fundamentally limiting their reliability. To overcome this limitation, affine frequency division multiplexing (AFDM) has emerged as a promising next-generation waveform \cite{bemani2023affine, sui2026generalized}, which leverages the discrete affine Fourier transform (DAFT) to achieve a complete delay-Doppler channel representation with superior resilience against mobility-induced impairments \cite{rou2024otfs, tao2025im}.
	
	Despite these theoretical advantages, the practical application of AFDM system at high carrier frequencies (e.g. mmWave) is mainly limited by transceiver hardware imperfections, especially oscillator phase noise (PN) \cite{piazzo2002analysis, sui2026generalized}. Caused by local oscillator instability, PN creates random phase changes that destroy subcarrier orthogonality and cause severe multiplicative distortion to the received signal. More importantly, this dynamic phase error strongly couples with the continuous delay-Doppler multipath parameters, creating a complex bilinear inverse problem \cite{wu2004ofdm}. As a result, traditional channel estimators become unreliable, which directly degrades accurate channel information and data recovery \cite{ wang2016channel}.
	
	Accurate channel state information is fundamental to reliable AFDM signal detection. Sparse Bayesian learning (SBL) provides an effective estimation paradigm by formulating doubly-dispersive channel recovery as a probabilistic inference problem. This intrinsically captures multipath sparsity without requiring predefined sparsity levels \cite{malioutov2005sparse, zheng2025channel}. However, standard SBL restricts continuous delay and Doppler shifts onto a fixed discrete dictionary, causing off-grid basis mismatches that severely degrade estimation accuracy \cite{wei2022offgrid}. The first-order approximation methods suffer from truncation errors under high mobility, leading to premature performance saturation. Alternatively, global grid densification expands the dictionary dimension, incurring a high computational burden that renders it impractical for real-time implementations \cite{luo2026joint, li2025grid}.
	
	Moreover, existing SBL estimators usually assume ideal hardware and separate channel estimation from PN tracking. In practical high-frequency AFDM systems, however, off-grid errors and time-varying PN are highly coupled, causing severe nonlinear distortions to the sensing dictionary. Traditional decoupled methods ignore this interaction, leading to severe error propagation between the two processes and resulting in an irreducible high-SNR error floor \cite{lin2007variational, mehrpouyan2012joint}. Therefore, a joint framework is required to simultaneously estimate the nonlinear PN and recover continuous channel parameters with low computational complexity.
	
	To address these challenges, we propose a joint PN and channel estimation method based on sparse Bayesian learning (JPNCE-SBL) for AFDM systems over doubly-dispersive channels. The joint estimation task is formulated as a bilinear sparse recovery problem, where an EM-based SBL framework is developed to iteratively decouple PN tracking from sparse channel recovery, thereby preventing error propagation. Specifically, a reduced-rank subspace projection is introduced to capture the dominant energy of the Wiener PN process, while a dynamic grid evolution strategy iteratively eliminates off-grid errors without global grid densification. Both components are integrated into a unified EM framework, where the channel and PN estimates are jointly updated at each iteration. Simulation results confirm that JPNCE-SBL significantly outperforms existing benchmarks in both NMSE and BER, closely approaching the perfect CSI bound under practical PN conditions. 
	\section{System Model}\label{sec:model} 
	Consider an $N$-subcarrier AFDM system with duration $T$ and bandwidth $B = N \Delta f$, where $\Delta f = 1/T$. The data symbols $\mathbf{x} = [x_{0}, \dots, x_{N-1}]^{\mathrm{T}} \in \mathbb{A}^{N \times 1}$ in the discrete affine Fourier (DAF) domain are transformed into the time domain via the inverse DAFT (IDAFT) \cite{bemani2023affine}:
	\begin{equation}
		s_{n} = \frac{1}{\sqrt{N}} \sum_{m=0}^{N-1} e^{j 2 \pi \left( c_{1} n^{2} + c_{2} m^{2} + \frac{1}{N} n m \right)} x_{m},
	\end{equation}
	where $c_{1}$ and $c_{2}$ denote the chirp parameters. In vector form, the above expression yields $\mathbf{s} = \mathbf{A}_{\mathrm{af}}^{\mathrm{H}} \mathbf{x}$, where $\mathbf{A}_{\mathrm{af}}^{\mathrm{H}} = \boldsymbol{\Lambda}_{c_{1}}^{\mathrm{H}} \mathbf{F}_{N}^{\mathrm{H}} \boldsymbol{\Lambda}_{c_{2}}^{\mathrm{H}}$ and $\boldsymbol{\Lambda}_{c_{i}} \triangleq \operatorname{diag} [e^{-j 2 \pi c_{i} 0^{2}}, \dots, e^{-j 2 \pi c_{i}(N-1)^{2}}]$. A chirp-periodic prefix (CPP) of length $N_{\mathrm{cpp}}$ is appended to $\mathbf{s}$ to mitigate inter-symbol interferencefrom multipath propagation.
	
	After propagating through a time-varying doubly-dispersive channel and being corrupted by receiver-side phase noise\cite{wang2016channel}, the received time-domain signal is modeled as
	\begin{equation}
		r_{n} = e^{j \phi_{n}} \sum_{p=1}^{P} h_{p} e^{j 2 \pi \nu_{p} n T_{s}} s_{[n-\ell_{p}]_{N}} + w_{n},
	\end{equation}
	where $e^{j \phi_{n}}$ denotes the $n$-th PN sample, $w_{n} \sim \mathcal{CN}(0, \sigma_{w}^{2})$ is the additive white Gaussian noise (AWGN), and $\{h_{p}, \ell_{p}, \nu_{p}\}$ denote the complex gain, normalized delay, and normalized Doppler shift of the $p$-th propagation path, respectively. The corresponding vector-form received signal is expressed as
	\begin{equation}
		\mathbf{r} = \mathbf{P} \sum_{p=1}^{P} h_{p} \boldsymbol{V}^{\nu_{p}} \boldsymbol{\Pi}^{\ell_{p}} \mathbf{s} + \mathbf{w},
	\end{equation}
	where $\mathbf{P} = \operatorname{diag}(e^{j \phi_{0}}, \dots, e^{j \phi_{N-1}})$, while $\boldsymbol{V}^{\nu_{p}}$ and $\boldsymbol{\Pi}^{\ell_{p}}$ denote the diagonal Doppler shift matrix and forward cyclic shift matrix, respectively. $\mathbf{w}$ is the noise vector.
	
	By performing the DAFT $\mathbf{y}_{\mathrm{af}} = \mathbf{A}_{\mathrm{af}} \mathbf{r}$ at the receiver, the DAF-domain received symbol $y_{\mathrm{af}}[\tilde{m}]$ can be expressed as
	\begin{equation}
		y_{\mathrm{af}}[\tilde{m}] = \sum_{m=0}^{N-1} H_{\mathrm{pn}}[\tilde{m}, m] x_{m} + \tilde{w}_{\tilde{m}},
	\end{equation}
	where the equivalent channel kernel $H_{\mathrm{pn}}[\tilde{m}, m]$ is given by
	\begin{equation}
		H_{\mathrm{pn}}[\tilde{m}, m] = \frac{1}{N} \sum_{p=1}^{P} h_{p} e^{j 2 \pi \left( c_{1} \ell_{p}^{2} - \frac{m \ell_{p}}{N} + c_{2} (m^{2} - \tilde{m}^{2}) \right)} \mathcal{G}_{p}(\tilde{m}, m).
	\end{equation}
	Here, the interaction function $\mathcal{G}_{p}(\tilde{m}, m) = \sum_{n=0}^{N-1} e^{j \phi_{n}} e^{-j \frac{2\pi}{N} (\tilde{m} - m + 2Nc_{1}\ell_{p} + \nu_{p})n}$ encapsulates the joint impact of PN and doubly-dispersive fading. It is worth noting that the PN term $e^{j\phi_{n}}$ disrupts the Dirichlet kernel structure inherent to the AFDM system, thereby inducing severe inter-carrier interference \cite{wu2004ofdm}. This disruption necessitates a robust framework to jointly estimate the sparse channel parameters and the PN parameter.
	
	In practical oscillators, the time-domain phase noise (PN) is widely described as a non-stationary Wiener process \cite{mathecken2011performance}. The phase evolution is given by $\phi_n = \phi_{n-1} + \delta_n$, where the innovation $\delta_n \sim \mathcal{N}(0, \sigma_{PN}^2)$. Assuming $\phi_{-1}=0$, the PN phase vector $\boldsymbol{\phi} = [\phi_{0}, \dots, \phi_{N-1}]^{\mathrm{T}}$ follows a multivariate Gaussian distribution, i.e., $\boldsymbol{\phi} \sim \mathcal{N}(\mathbf{0}, \mathbf{R}_{\phi})$. The elements of the covariance matrix $\mathbf{R}_{\phi} \in \mathbb{R}^{N \times N}$ are compactly defined as $[\mathbf{R}_{\phi}]_{m,n} = \sigma_{PN}^2 (\min(m,n)+1)$.
	
	Furthermore, to alleviate the high-dimensional estimation burden, $\boldsymbol{\phi}$ is projected onto a lower-dimensional subspace via the singular value decomposition $\mathbf{R}_{\phi} = \mathbf{U} \mathbf{D} \mathbf{U}^{\mathrm{T}}$, yielding a basis expansion model:
	\begin{equation}
		\boldsymbol{\phi} \approx \mathbf{B} \boldsymbol{\eta},
	\end{equation}
	where $\mathbf{B} = \mathbf{U}_{1:L} \mathbf{D}_{1:L}^{1/2} \in \mathbb{R}^{N \times L}$ comprises the $L$ dominant basis vectors with $L \ll N$, and $\boldsymbol{\eta} \sim \mathcal{N}(\mathbf{0}, \mathbf{I}_L)$ denotes the reduced-rank PN coefficient vector. The parameterized time-domain PN matrix is then given by $\mathbf{P}(\boldsymbol{\eta}) = \operatorname{diag}(e^{j \mathbf{B}\boldsymbol{\eta}})$. In the proposed JPNCE-SBL framework, tracking the low-dimensional $\boldsymbol{\eta}$ instead of $\boldsymbol{\phi}$ significantly reduces computational overhead while ensuring numerical stability.

	Next, we consider the joint channel estimation problem. To exploit the inherent sparsity of the doubly-dispersive channel, the estimation task is formulated as a sparse recovery problem by establishing a virtual sampling grid \cite{malioutov2005sparse, wei2022offgrid}.  
	The delay and Doppler domains are discretized within the ranges $[0, \ell_{\max}]$ and $[-k_{\max}-1, k_{\max}+1]$ with resolutions $r_\tau = \frac{\ell_{\max}}{M_\tau-1}$ and $r_\nu = \frac{2k_{\max}+2}{M_\nu-1}$, respectively. The resulting virtual sampling delay and Doppler vectors are $ [\tilde{\ell}_0, \dots, \tilde{\ell}_{M_\tau-1}]^{\mathrm{T}}$ and $ [\tilde{k}_0, \dots, \tilde{k}_{M_\nu-1}]^{\mathrm{T}}$, where $\tilde{\ell}_a = ar_\tau$ and $\tilde{k}_b = br_\nu - k_{\max} - 1$. The sampling grid is then defined as $\overline{\mathbf{G}}= \{\bar{g}_m\}_{m=0}^{M_S-1}=\{\bar{\ell}_m, \bar{k}_m\}_{m=0}^{M_S-1}$ with $M_S = M_\tau M_\nu$, comprising all combinations of virtual parameters where $\bar{\ell}_m = \tilde{\ell}_a$ and $\bar{k}_m = \tilde{k}_b$ for $a = \lfloor m/M_\nu \rfloor$ and $b = m \mod{M_\nu}$.  Then, by mapping the physical parameters onto this grid,  the received signal can
	be reformulated as
	\begin{equation}\label{eq:signal_model_grid}
		\mathbf{r} = \mathbf{P}(\boldsymbol{\eta}) \boldsymbol{\Phi}(\overline{\boldsymbol{G}}) \overline{\mathbf{h}} + \mathbf{w},
	\end{equation}
	where $\overline{\mathbf{h}} \in \mathbb{C}^{M_{S} \times 1}$ is the sparse channel vector to be estimated,  the dictionary matrix $\boldsymbol{\Phi}(\overline{\boldsymbol{G}}) \in \mathbb{C}^{N \times M_{S}}$ consists of the steering vectors $\boldsymbol{\phi}(\bar{g}_{m}) = \boldsymbol{V}^{\bar{\nu}_{m}} \boldsymbol{\Pi}^{\bar{\ell}_{m}} \mathbf{s}$ for all grid points $m = 0, \dots, M_{S}-1$, and the diagonal matrix $\mathbf{P}(\boldsymbol{\eta}) \in \mathbb{C}^{N \times N}$ is the  PN error parameterized by the coefficient vector $\boldsymbol{\eta}$. The joint estimation of $\overline{\mathbf{h}}$ and $\boldsymbol{\eta}$ is challenging due to their bilinear coupling in \eqref{eq:signal_model_grid}, the non-convexity induced by $\mathbf{P}(\boldsymbol{\eta}) = \operatorname{diag}(e^{j\mathbf{B}\boldsymbol{\eta}})$, and the dictionary mismatch caused by off-grid effects. To address these challenges, we propose the JPNCE-SBL framework in the following section.
	
	\section{The Proposed JPNCE-SBL Method for AFDM}\label{sec:method}  
	In this section, we develop a joint channel and PN estimation algorithm based on the SBL framework \cite{li2025grid}. To exploit the inherent sparsity of the channel vector $\overline{\mathbf{h}}$, a hierarchical Gaussian-Gamma prior is adopted \cite{wei2022offgrid}, where the first layer assumes that the entries of $\overline{\mathbf{h}}$ follow a conditional complex Gaussian distribution: 
	\begin{equation}
		\label{eq:h_prior}
		p(\overline{\mathbf{h}} | \boldsymbol{\gamma}) = \mathcal{CN}(\overline{\mathbf{h}} | \mathbf{0}, \boldsymbol{\Gamma}),
	\end{equation}
	\begin{figure}[th]
		\centering
		\includegraphics[width=0.85\linewidth]{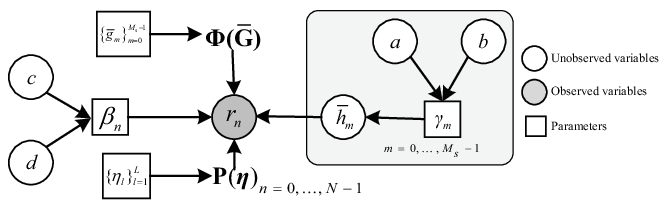}
		\caption{\small{The graphical model for the JPNCE-SBL method.}}
		\label{fig1} 
	\end{figure}  
	where $\boldsymbol{\Gamma} = \operatorname{diag}(\boldsymbol{\gamma})$ and $\boldsymbol{\gamma} = [\gamma_{0}, \dots, \gamma_{M_{S}-1}]^{\mathrm{T}}$. In the second layer, to ensure non-informative priors, each hyperparameter $\gamma_{m}$ and the noise precision $\beta = \sigma_{w}^{-2}$ are independently governed by Gamma distributions with small constants: 
	\begin{equation}
		p(\boldsymbol{\gamma} | a, b) = \prod_{m=0}^{M_{S}-1} \Gamma(\gamma_{m} | a, b),   p(\mathbf{w} | \beta) = \mathcal{CN} (\mathbf{w} | \mathbf{0}, \beta^{-1} \mathbf{I}_{N} ).
	\end{equation}The corresponding graphical model for the joint estimation process is illustrated in Fig.~\ref{fig1}. It is worth noting that marginalizing over $\boldsymbol{\gamma}$ naturally yields a sparsity-enforcing Student-$t$ marginal prior for each entry of $\overline{\mathbf{h}}$ \cite{malioutov2005sparse}. Based on the signal model established in Section~\ref{sec:model}, the conditional likelihood of the observation is given by 
	\begin{equation}
		\label{eq:likelihood}
		p(\mathbf{r} | \overline{\mathbf{h}}; \beta, \overline{\mathbf{G}},\boldsymbol{\eta}) = \mathcal{CN}(\mathbf{r} | \mathbf{P}(\boldsymbol{\eta})\boldsymbol{\Phi}(\overline{\mathbf{G}}) \overline{\mathbf{h}},  \beta^{-1} \mathbf{I}_{N}),
	\end{equation}
	Thus, let the parameter set $\boldsymbol{\theta} = \{\boldsymbol{\gamma}, \beta,\overline{\mathbf{G}}, \boldsymbol{\eta}\}$, the conditional posterior follows $p(\overline{\mathbf{h}} | \mathbf{r}; \boldsymbol{\theta}) \propto p(\mathbf{r}| \overline{\boldsymbol{h}} ; \beta,\overline{\mathbf{G}}, \boldsymbol{\eta}) p(\overline{\boldsymbol{h}} | \gamma) = \mathcal{CN}(\overline{\mathbf{h}} | \boldsymbol{\mu}, \boldsymbol{\Sigma})$, where the posterior statistics $\boldsymbol{\mu}$ and $\boldsymbol{\Sigma}$\cite{li2025grid}:
	\begin{equation}
		\label{eq:posterior_stat}
		\boldsymbol{\mu} = \beta \boldsymbol{\Sigma} \boldsymbol{\Psi}^{\mathrm{H}} \mathbf{r},\quad
		\boldsymbol{\Sigma} = \boldsymbol{\Gamma} - \boldsymbol{\Gamma} \boldsymbol{\Psi}^{\mathrm{H}} \left( \beta^{-1} \mathbf{I}_{N} + \boldsymbol{\Psi} \boldsymbol{\Gamma} \boldsymbol{\Psi}^{\mathrm{H}} \right)^{-1} \boldsymbol{\Psi} \boldsymbol{\Gamma}, 
	\end{equation}
	where $\boldsymbol{\Psi} = \mathbf{P}(\boldsymbol{\eta}) \boldsymbol{\Phi}(\overline{\mathbf{G}})$ denotes the effective sensing matrix. Accoring to the maximum a posteriori criterion, the sparse channel estimate is directly obtained as $\hat{\overline{\mathbf{h}}} = \boldsymbol{\mu}$.
	\subsection{The Proposed JPNCE-SBL} 
	Since the posterior statistics depend heavily on the unknown parameters in $\boldsymbol{\theta}$, the Expectation-Maximization (EM) algorithm is employed for the iterative estimation of $\boldsymbol{\theta}$. Specifically, the marginal log-likelihood is maximized via the evidence lower bound \cite{zhao2020sparse}. During the E-step of the $(i+1)$-th iteration, by taking the expectation of the joint log-likelihood $\ln p(\mathbf{r}, \overline{\mathbf{h}}; \boldsymbol{\theta})$ with respect to $p(\overline{\mathbf{h}} | \mathbf{r}; \boldsymbol{\theta}^{(i)})$, the objective log-likelihood function $\mathcal{Q}(\boldsymbol{\theta}, \boldsymbol{\theta}^{(i)})$ can be derived as:
	\begin{equation}
		\label{eq:Q_function_expanded}
		\begin{aligned}
			\mathcal{Q}(\boldsymbol{\theta}, \boldsymbol{\theta}^{(i)})&=\langle\ln p({\mathbf{r}}, \overline{\mathbf{h}} ;  {\boldsymbol{\theta}})\rangle_{p\left(\overline{\mathbf{h}} |  {\mathbf{r}} ; \boldsymbol{\theta}^{(i)}\right)}  \\ 
			&\propto N \ln \beta - \beta \left\| \mathbf{r} - \boldsymbol{\Psi} \boldsymbol{\mu}^{(i)} \right\|_{2}^{2} - \beta \operatorname{Tr} \left( \boldsymbol{\Psi} \boldsymbol{\Sigma}^{(i)} \boldsymbol{\Psi}^{\mathrm{H}} \right) \\
			& + \sum_{m=0}^{M_{S}-1} \left( \ln \gamma_{m}^{-1} - \gamma_{m}^{-1} \Xi_m^{(i)} \right),
		\end{aligned}
	\end{equation}
	where $\Xi_m^{(i)} = \Sigma_{m,m}^{(i)} + |\mu_{m}^{(i)}|^{2}$,  and  the second equality exploits the expectation trace property and $\langle  \overline{\mathbf{h}}\overline{\mathbf{h}}^{\mathrm{H}}\rangle_{p\left(\overline{\mathbf{h}} |  {\mathbf{r}} ; \boldsymbol{\theta}^{(i)}\right)}= \boldsymbol{\Sigma}^{(i)} + \boldsymbol{\mu}^{(i)}(\boldsymbol{\mu}^{(i)})^{\mathrm{H}}$. Subsequently, in the M-step, the parameters are updated via $\boldsymbol{\theta}^{(i+1)} = \arg\max_{\boldsymbol{\theta}}\, \mathcal{Q}(\boldsymbol{\theta}, \boldsymbol{\theta}^{(i)})$\cite{liu2012efficient}. By exploiting the separable structure of $\mathcal{Q}$ with respect to $\boldsymbol{\gamma}$, $\beta$, $\overline{\mathbf{G}}$, and $\boldsymbol{\eta}$, the optimization process is effectively decoupled and executed across three sequential stages.
	
	\subsubsection{Stage 1: Updating Hyperparameters $\boldsymbol{\gamma}^{(i+1)}$ and $\beta^{(i+1)}$} 
	When given the conditional posterior $p(\overline{\mathbf{h}} | \mathbf{r}; \boldsymbol{\theta}^{(i)})$, the hyperparameters $\boldsymbol{\gamma}^{(i+1)}$ and $\beta^{(i+1)}$ are updated by maximizing the objective function $\mathcal{Q}(\boldsymbol{\theta}, \boldsymbol{\theta}^{(i)})$ with respect to $\boldsymbol{\gamma}$ and $\beta$, i.e., $(\boldsymbol{\gamma}^{(i+1)},\beta^{(i+1)})=\arg\max_{\boldsymbol{\gamma},\beta}\mathcal{Q}(\boldsymbol{\theta}, \boldsymbol{\theta}^{(i)})$. Then, by setting the partial derivatives with respect to $\gamma_{m}$ and $\beta$ to zero, respectively, the closed-form update rules can be  obtained as 
	\begin{small} 
		\begin{equation}
			\label{eq:gamma_beta_update}
			\beta^{(i+1)} = \frac{N + c - 1}{d + \mathcal{C}_{\beta}^{(i)}},
			\gamma_{m}^{(i+1)} = \frac{\sqrt{ 4\rho\left(|\mu_{m}^{(i)}|^{2} + \Sigma_{m,m}^{(i)}+1 \right)} - 1}{2\rho},
		\end{equation}
	\end{small}where the residual term  $\mathcal{C}_{\beta}^{(i)} =  \| \mathbf{r} - \boldsymbol{\Psi}^{(i)} \boldsymbol{\mu}^{(i)}  \|_{2}^{2} +  (\beta^{(i)} )^{-1} \sum_{m=0}^{M_{S}-1}  ( 1 - {\Sigma_{m,m}^{(i)}}/{\gamma_{m}^{(i)}} ) $. When updating $\boldsymbol{\gamma}^{(i+1)}$, we extract the $\hat{P}$ indices with the largest magnitudes to form the active support set $\mathcal{P}$. The corresponding delay and Doppler components are preliminarily estimated as $\bar{g}_{p}^{(i)} = \{\bar{\ell}_{p}^{(i)}, \bar{\nu}_{p}^{(i)}\}$ for $p \in \mathcal{P}$.  
	It is noted that conventional SBL methods cannot handle PN-induced dictionary mismatches and off-grid effects. While Off Grid SBL method uses first-order linear approximations, it remains sensitive to modeling errors. To address grid limitations,  we resolve these issues in Stage 3 by evolving the grid $\overline{\mathbf{G}}$ to match the continuous parameters, which also ensure precise PN estimation.
	
	\subsubsection{Stage 2: Updating PN Coefficient Vector $\boldsymbol{\eta}^{(i+1)}$}
	
	In this stage, we update the reduced-rank PN coefficients $\boldsymbol{\eta}^{(i+1)}$ using the dictionary $\boldsymbol{\Phi}(\overline{\mathbf{G}}^{(i)})$ and the channel mean $\boldsymbol{\mu}^{(i)}$ from the previous iteration. The objective is to minimize the cost function $\mathcal{L}(\boldsymbol{\eta})$ while fixing the current sampling grid $\overline{\mathbf{G}}^{(i)}$:
	\begin{equation} \label{eq:cost_function_raw}
		\hat{\boldsymbol{\eta}}^{(i+1)} = \underset{\boldsymbol{\eta}}{\arg \min } \,\, \mathcal{L}(\boldsymbol{\eta}^{(i)} ),
	\end{equation}
	where $\mathcal{L}(\boldsymbol{\eta}) = \log \operatorname{det}(\boldsymbol{\Sigma}) + \big(\mathbf{r} - \boldsymbol{\mu}_{\mathbf{r}}(\boldsymbol{\eta}, \overline{\mathbf{G}}^{(i)})\big)^{\mathrm{H}} \boldsymbol{\Sigma}^{-1} \big(\mathbf{r} - \boldsymbol{\mu}_{\mathbf{r}}(\boldsymbol{\eta}, \overline{\mathbf{G}}^{(i)})\big) + {1}/{2} \boldsymbol{\eta}^{\mathrm{T}} \boldsymbol{\eta}$, and the reconstructed signal is $\boldsymbol{\mu}_{\mathbf{r}}(\boldsymbol{\eta}, \overline{\mathbf{G}}^{(i)}) = \mathbf{P}(\boldsymbol{\eta}) \boldsymbol{\Phi}(\overline{\mathbf{G}}^{(i)}) \boldsymbol{\mu}^{(i)}$.   
	To obtain a closed-form solution for this non-convex problem, we perform two steps. First, we freeze the noise covariance $\boldsymbol{\Sigma}$ using the estimate from the $i$-th iteration. Second, since the PN innovation variance of practical oscillators is usually small, we can apply the small-angle approximation like Taylor series expansion to the PN matrix $\mathbf{P}(\boldsymbol{\eta})$, i.e.,
	\begin{equation} \label{eq:small_angle_approx}
		\mathbf{P}(\boldsymbol{\eta}) = \operatorname{diag}\left( e^{j \mathbf{B} \boldsymbol{\eta}} \right) \approx \mathbf{I}_N + j \operatorname{diag}(\mathbf{B} \boldsymbol{\eta}).
	\end{equation}
	
	Then, the object function $\mathcal{L}(\boldsymbol{\eta})$ can be reformulated into a regularized quadratic form:
	\begin{small}
		\begin{equation} \label{eq:quadratic_cost}
			\begin{aligned}
				\mathcal{L}(\boldsymbol{\eta}) &\approx (\overline{\mathbf{r}}^{(i)} - \mathbf{V}^{(i)} \boldsymbol{\eta} )^{\mathrm{H}} \boldsymbol{\Sigma}^{-1} (\overline{\mathbf{r}}^{(i)} - \mathbf{V}^{(i)} \boldsymbol{\eta} ) + \frac{1}{2} \boldsymbol{\eta}^{\mathrm{T}} \boldsymbol{\eta},\\
				& \approx \boldsymbol{\eta}^{\mathrm{T}} {\Re} ( \mathbf{V}^{(i)\mathrm{H}} \boldsymbol{\Sigma}^{-1} \mathbf{V}^{(i)} ) \boldsymbol{\eta} - 2 {\Re} ( \overline{\mathbf{r}}^{(i)\mathrm{H}} \boldsymbol{\Sigma}^{-1} \mathbf{V}^{(i)} ) \boldsymbol{\eta} + \frac{1}{2} \boldsymbol{\eta}^{\mathrm{T}} \boldsymbol{\eta},
			\end{aligned}
		\end{equation}
	\end{small}where $\overline{\mathbf{r}}^{(i)}$ is the error relative to observation and the signal without phase noise, and $\mathbf{V}^{(i)}$ is coupling matrix, given by
	\begin{equation} \label{eq:auxiliary_vars}
		\begin{aligned}
			\overline{\mathbf{r}}^{(i)} &= \mathbf{r} - \boldsymbol{\Phi}(\overline{\mathbf{G}}^{(i)}) \boldsymbol{\mu}^{(i)}, \\
			\mathbf{V}^{(i)} &= j \operatorname{diag}( \boldsymbol{\Phi}(\overline{\mathbf{G}}^{(i)}) \boldsymbol{\mu}^{(i)} ) \mathbf{B}.
		\end{aligned}
	\end{equation} 
	
	Finally, we set the gradient of \eqref{eq:quadratic_cost} with respect to $\boldsymbol{\eta}$ to zero. This leads to the closed-form update rule for the PN coefficients, $\boldsymbol{\eta}^{(i+1)} $, which can be given as
	\begin{small}
		\begin{equation} \label{eq:final_eta_update}
			\boldsymbol{\eta}^{(i+1)} =  [ {\Re} (( \mathbf{V}^{(i)})^\mathrm{H} \boldsymbol{\Sigma}^{-1} \mathbf{V}^{(i)} ) + \frac{1}{2}\mathbf{I}_L  ]^{-1} {\Re} ( ( \mathbf{V}^{(i)})^\mathrm{H}\boldsymbol{\Sigma}^{-1} \overline{\mathbf{r}}^{(i)} ).
		\end{equation}
	\end{small} 
	After computing $\boldsymbol{\eta}^{(i+1)}$, we reconstruct the full-dimensional PN vector as $\hat{\boldsymbol{\phi}}^{(i+1)} = \mathbf{B} \boldsymbol{\eta}^{(i+1)}$. The PN matrix $\mathbf{P}(\boldsymbol{\eta}^{(i+1)})$ is then refreshed for the subsequent channel estimation and grid refinement stages.
	\subsubsection{Stage 3: Updating Doppler Grid via Grid Evolution} 
	We develop a dynamic grid evolution (GE) strategy to mitigate the discretization errors of fixed grids and the approximation errors in conventional SBL. This method iteratively refines the virtual Doppler grid points $\bar{k}_{m}$ without expanding the measurement matrix. Thus, the refined points converge toward the true continuous Doppler parameters to improve estimation accuracy. Specifically, we approximate the dictionary matrix $\boldsymbol{\Phi}(\overline{\mathbf{G}})$ via a first-order linear expansion with respect to $\bar{k}_{m}$ around the current grid points: 
	\begin{equation}
		\label{eq:dict_taylor}
		\hat{\boldsymbol{\Phi}}(\overline{\mathbf{G}}) \approx \boldsymbol{\Phi}(\overline{\mathbf{G}}) + \boldsymbol{\Omega}(\overline{\mathbf{G}}) \operatorname{diag}(\boldsymbol{\xi}),
	\end{equation}
	where $\boldsymbol{\xi} = [\xi_0, \dots, \xi_{M_S-1}]^{\mathrm{T}}$ is the off-grid Doppler residual vector, and the derivative matrix $\boldsymbol{\Omega}(\overline{\mathbf{G}})$ is constructed with columns defined as $\boldsymbol{\omega}(\bar{g}_m) = {\partial\boldsymbol{\phi}(\bar{g}_m)}/{\partial \bar{k}_m}$.
	
	In the M-step, we update the grid perturbation $\boldsymbol{\xi}$ by minimizing the expected residual error. We utilize the posterior mean $\boldsymbol{\mu}^{(i)}$ and covariance $\boldsymbol{\Sigma}^{(i)}$ from the E-step to perform this update. Then, the optimization problem can be shown as:
	\begin{small}  
		\begin{equation}
			\label{eq:stage3_opt_obj}
			\begin{aligned}
				\boldsymbol{\xi}^{(i+1)} &= \arg\min_{\boldsymbol{\xi}}\;  \mathcal{Q}^{({i})}(\boldsymbol{\theta}, \boldsymbol{\theta}^{(i)})\\
				& =   \arg\min_{\boldsymbol{\xi}}  \{ \big\| \mathbf{r} - \mathbf{P}(\boldsymbol{\eta}^{(i+1)})\hat{\boldsymbol{\Phi}}({\overline{\mathbf{G}}^{(i)}})  \boldsymbol{\mu}^{(i)} \big\|_2^2  \\
				&   + \operatorname{Tr} \Big[\mathbf{P}(\boldsymbol{\eta}^{(i+1)})\hat{\boldsymbol{\Phi}}({\overline{\mathbf{G}}^{(i)}})  \boldsymbol{\Sigma}^{(i)}(\mathbf{P}(\boldsymbol{\eta}^{(i+1)})\hat{\boldsymbol{\Phi}}({\overline{\mathbf{G}}^{(i)}}) )^{H}\Big] \}.
			\end{aligned}
		\end{equation}
	\end{small} Then, according to the properties of trace operator and the Hadamard product, the objective function in \eqref{eq:stage3_opt_obj} can be reduced to a standard quadratic form and shown as 
	\begin{equation}
		\label{eq:quadratic_form}
		\mathcal{L}(\boldsymbol{\xi}) \approx \boldsymbol{\xi}^{\mathrm{T}} \mathbf{Q}^{(i)} \boldsymbol{\xi} - 2\mathbf{d}^{(i)\mathrm{T}} \boldsymbol{\xi} + C,
	\end{equation}
	where $C$ is a constant term independent of $\boldsymbol{\xi}$, and the matrix $\mathbf{Q}^{(i)}$ and the vector $\mathbf{d}^{(i)}$ can be defined as 
	\begin{small}
		\begin{equation}
			\label{eq:Q_d_def}
			\begin{aligned}
				\mathbf{Q}^{(i)} &= \Re \Big\{ \big( (\boldsymbol{\Omega}^{(i)})^{\mathrm{H}} \boldsymbol{\Omega}^{(i)} \big)^* \odot \big( \boldsymbol{\mu}^{(i)}(\boldsymbol{\mu}^{(i)})^{\mathrm{H}} + \boldsymbol{\Sigma}^{(i)} \big) \Big\}, \\
				\mathbf{d}^{(i)} &= \Re \Big\{ \operatorname{diag}\big( \boldsymbol{\mu}^{(i)*} \big) (\boldsymbol{\Omega}^{(i)})^{\mathrm{H}}  (\mathbf{P}^{(i+1)})^{\mathrm{H}}  \big( \mathbf{r} - \mathbf{P}^{(i+1)}\boldsymbol{\Phi}^{(i)} \boldsymbol{\mu}^{(i)} \big) \\
				&\quad \quad \quad - \operatorname{diag} \big( (\boldsymbol{\Omega}^{(i)})^{\mathrm{H}} \boldsymbol{\Phi}^{(i)} \boldsymbol{\Sigma}^{(i)} \big) \Big\}.
			\end{aligned}
		\end{equation}
	\end{small}To reduce complexity, we update the grid only on the active set $\mathcal{P}$ from Stage 1. Specifically, let $\boldsymbol{\xi}_{\mathcal{P}}$ be the doppler offset vector for $\mathcal{P}$. We extract the sub-matrix $\mathbf{Q}_{\mathcal{P}}^{(i)}$ and sub-vector $\mathbf{d}_{\mathcal{P}}^{(i)}$. Then, we can obtain the closed-form solution:
	\begin{equation}
		\label{eq:xi_update_final}
		\boldsymbol{\xi}_{\mathcal{P}}^{(i+1)} = \left(\mathbf{Q}_{\mathcal{P}}^{(i)}\right)^{-1} \mathbf{d}_{\mathcal{P}}^{(i)}.
	\end{equation} 
	Moreover,  to maintain grid integrity and avoid overlapping, we constrain the offsets to the resolution interval, and the doppler grid offset can be given as
	\begin{equation}\label{eq:xi_clip}
		\xi_{p}^{(i+1)} = \operatorname{clip}(\xi_{p}^{(i+1)}, - {r_{\nu}}/{2},{r_{\nu}}/{2}), \quad \forall p \in \mathcal{P}.
	\end{equation}
	Finally, we update the virtual Doppler grid for the next iteration as $\bar{k}_{p}^{(i+1)} = \bar{k}_{p}^{(i)} + \xi_{p}^{(i+1)}$. Note that the JPNCE-SBL algorithm stops when the parameter change satisfies $\|\boldsymbol{\gamma}^{(i+1)} - \boldsymbol{\gamma}^{(i)}\|_2 / \|\boldsymbol{\gamma}^{(i)}\|_2 < \epsilon$ or reaches the max iterations $n_{\mathrm{iter}}$. After stopping, we find the active set $\hat{\mathcal{P}}$ by keeping indices with $\gamma_{p}^{(i+1)} > \varepsilon$, where $\varepsilon$ is a small threshold. Then, we count the number of paths as $\hat{P} = |\hat{\mathcal{P}}|$.  Finally, we get the channel gains, phase noise, the final delays, and the Doppler parameters.
	The complete JPNCE-SBL framework is summarized in Algorithm~\ref{alg_jpnce_sbl}.
	\begin{algorithm}[h]
		\caption{Proposed JPNCE-SBL Algorithm}
		\label{alg_jpnce_sbl}
		\renewcommand{\algorithmicrequire}{\textbf{Input:}}
		\renewcommand{\algorithmicensure}{\textbf{Initialize:}}
		\begin{algorithmic}[1]
			\REQUIRE  $\mathbf{r}$,  $\overline{\mathbf{G}}^{(0)}$,  $\mathbf{B}$,  $\epsilon$,   $n_{\mathrm{iter}}$
			\ENSURE  $i=0$,  $\boldsymbol{\eta}^{(0)}=\mathbf{0}$,  $\boldsymbol{\gamma}^{(0)}=\mathbf{1}$,  $\beta^{(0)}=1$,  $\boldsymbol{\xi}^{(0)}=\mathbf{0}$. Compute initial $\boldsymbol{\mu}^{(0)}$ and $\boldsymbol{\Sigma}^{(0)}$.
			\WHILE{$\|\boldsymbol{\gamma}^{(i)} - \boldsymbol{\gamma}^{(i-1)}\|_2 / \|\boldsymbol{\gamma}^{(i-1)}\|_2 \ge \epsilon$ \textbf{and} $i < n_{\mathrm{iter}}$}
			\STATE $i \leftarrow i+1$
			\STATE Update hyperparameters $\boldsymbol{\gamma}^{(i)}$ and $\beta^{(i)}$ via \eqref{eq:gamma_beta_update} using $\boldsymbol{\mu}^{(i-1)}$ and $\boldsymbol{\Sigma}^{(i-1)}$, then extract the active set $\mathcal{P}$.
			\STATE Update the reduced-rank PN coefficients $\boldsymbol{\eta}^{(i)}$ via \eqref{eq:final_eta_update} and reconstruct $\mathbf{P}(\boldsymbol{\eta}^{(i)})$.
			\STATE Compute and clip offsets $\boldsymbol{\xi}_{\mathcal{P}}^{(i)}$ via \eqref{eq:xi_update_final}, then update the virtual grid $\bar{k}_p^{(i)} = \bar{k}_p^{(i-1)} + \xi_p^{(i)}, \forall p \in \mathcal{P}$.
			\STATE Reconstruct the dictionary $\boldsymbol{\Phi}(\overline{\mathbf{G}}^{(i)})$ and evaluate the new posterior statistics $\boldsymbol{\mu}^{(i)}$, $\boldsymbol{\Sigma}^{(i)}$.
			\ENDWHILE 
			\RETURN $\hat{P} = |\hat{\mathcal{P}}|$, $\hat{\overline{\mathbf{h}}}$, $\hat{\boldsymbol{\eta}}$,  $\hat{\boldsymbol{\ell}}$,  $\hat{\boldsymbol{\nu}}$.	
		\end{algorithmic} 
	\end{algorithm}
	\subsection{Computational Complexity Analysis}  
	The overall complexity of JPNCE-SBL is asymptotically bounded by the E-step. The intermediate operations require much less computation. Specifically, hyperparameter updates, dictionary construction, and grid evolution only contribute lower-order terms such as $\mathcal{O}(NL^2)$ and $\mathcal{O}(|\mathcal{P}|^3)$ \cite{zhao2020sparse, mehrpouyan2012joint}. In the E-step, updating the posterior statistics inherently requires inverting an $M_S \times M_S$ matrix. Since the observation dimension is typically smaller than the high-resolution grid size ($N < M_S$), we invoke the Woodbury matrix identity to fundamentally reduce the inversion dimension from $M_S$ to $N$, lowering the E-step complexity to $\mathcal{O}(N^3 + N^2 M_S)$. 
	Both the phase noise parameters and the active paths exhibit high sparsity ($L \ll N$ and $|\mathcal{P}| \ll M_S$). As a result, the lower-order terms become negligible. Thus, the total complexity per iteration becomes $\mathcal{O}(N^3 + N^2 M_S)$. This means the proposed PN tracking and grid evolution introduce almost no extra overhead. Consequently, the algorithm achieves massive performance gains at the original complexity level, which ensures its real-time feasibility in practical high-mobility systems.
	\section{Simulation Results}\label{sec:sim}
	The proposed JPNCE-SBL algorithm is evaluated in an AFDM system comprising $N = 64$ subcarriers, with carrier frequency $f_c = 30~\text{GHz}$, subcarrier spacing $\Delta f = 15~\text{kHz}$, and $4$-QAM modulations. The doubly-dispersive channel consists of $P =3$ independent paths, following the Jakes model with a maximum normalized delay $\ell_{\max} =12$ and Doppler shift $k_{\max} = 3$. For reliable channel acquisition, five pilot symbols are multiplexed with a $30~\text{dB}$ power boost. The receiver-side PN is modeled as a discrete-time Wiener process \cite{mathecken2011performance}. For the SBL configuration, grid resolutions are $r_{\tau} = r_{\nu} = 1$, initialized with $\rho = 10^{-2}$, $c = d = 10^{-6}$, and $100$ maximum iterations. For data recovery, an MMSE detector is employed. The proposed scheme is benchmarked against Newton-OMP, OffGrid-SBL\cite{wei2022offgrid}, and Grid Fission SBL\cite{pote2023maximum} under two scenarios: an ideal oscillator (Without PN) and practical  PN impairments (With PN). Performance is quantified via normalized mean square error (NMSE) and bit error rate (BER).
	
	\begin{figure}[h]
		\centering
		\includegraphics[width=0.85\linewidth]{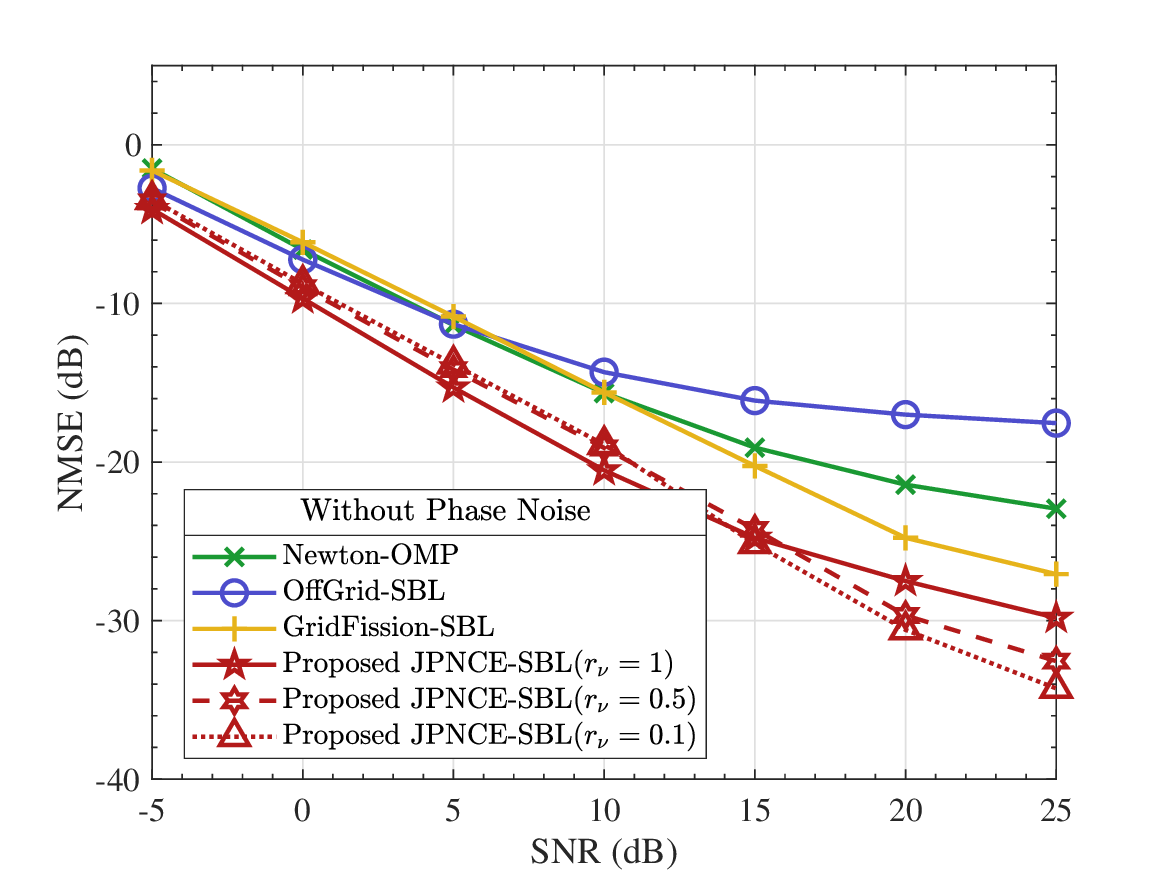}
		\caption{\small{NMSE performance comparison of different estimators versus SNR under an ideal oscillator scenario.}}
		\label{fig2}
		\vspace{-1em}
	\end{figure}	
	To evaluate the performance gains of the proposed grid evolution strategy, Fig.~\ref{fig2} presents the NMSE results under an ideal oscillator scenario. It can be observed that Newton-OMP and OffGrid-SBL suffer from severe high-SNR error floors due to discretization errors on fixed grids, while GridFission-SBL achieves the best baseline performance by reducing grid mismatch but still saturates at high SNR. In contrast, the proposed JPNCE-SBL consistently achieves superior accuracy by continuously refining the virtual grid, exhibiting no error floor across the entire SNR range. Furthermore, the NMSE performance of JPNCE-SBL remains stable across different grid resolutions $r_\nu \in \{0.1, 0.5, 1\}$, with only small performance loss as $r_\nu$ increases, demonstrating that the proposed grid evolution strategy effectively handles coarse grid initializations. This confirms the robustness of JPNCE-SBL against grid resolution variations and validates the effectiveness of the proposed dynamic grid update mechanism.
	\begin{figure}[htbp]
		\centering
		\includegraphics[width=0.83\linewidth]{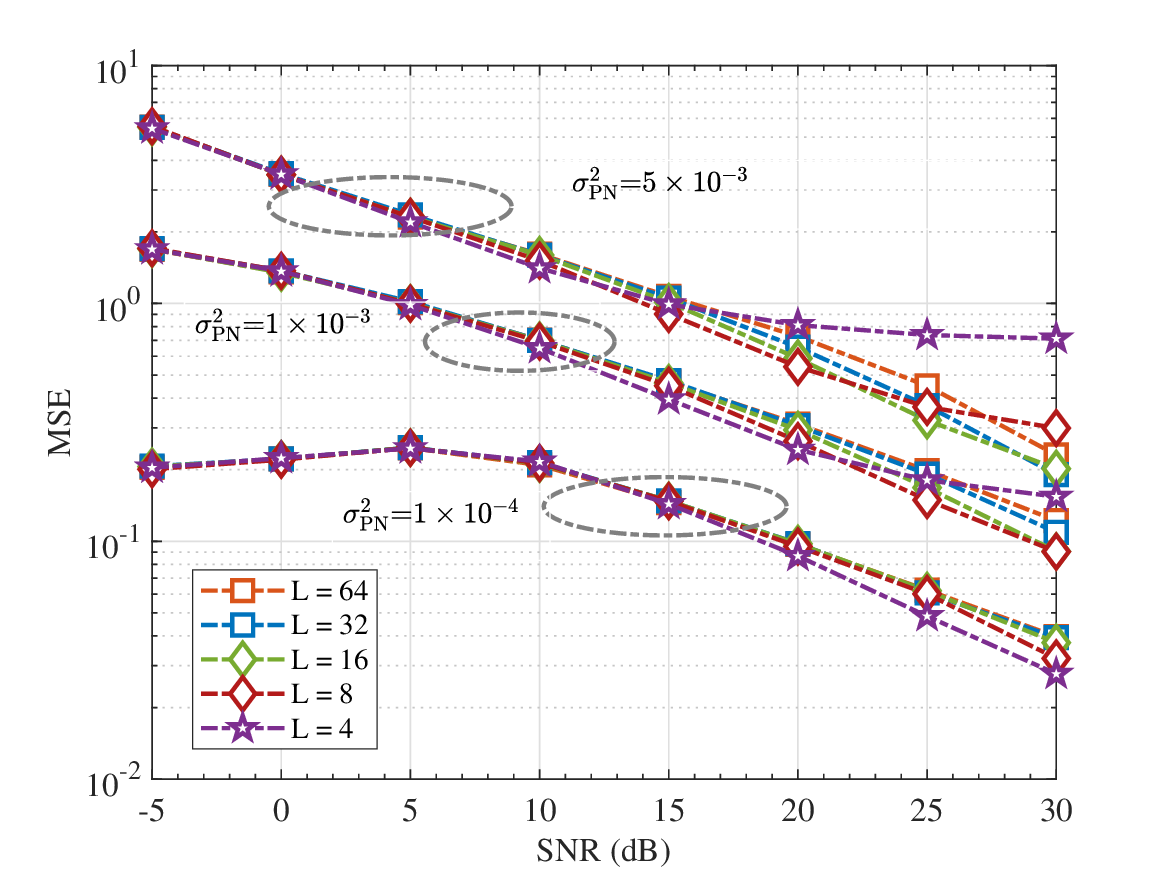}
		\caption{\small{MSE performance of the proposed subspace projection method for PN estimation under different subspace dimensions $L$ and PN innovation variances.}}
		\label{fig4}
		\vspace{-1em}
	\end{figure}
	
	\begin{figure}[htbp]
		\centering
		\includegraphics[width=0.83\linewidth]{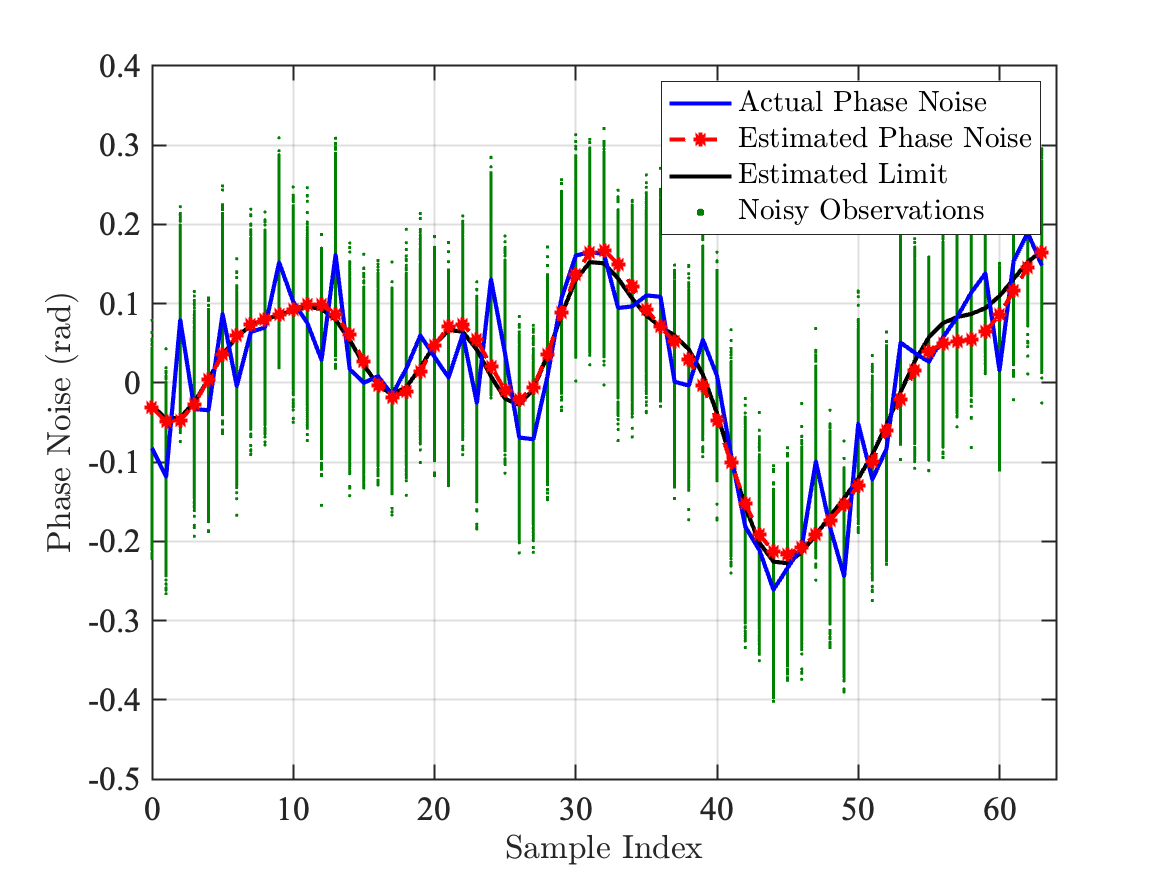}
		\caption{\small{PN trajectory tracking performance of the proposed subspace projection method.}}
		\label{fig5}
		\vspace{-2em}
	\end{figure} 
	Fig.~\ref{fig4} and Fig.~\ref{fig5} evaluate the proposed subspace projection method through the PN tracking MSE and temporal trajectory, respectively. From Fig.~\ref{fig4}, an irreducible error floor emerges at high SNRs due to the truncation loss of the finite-rank approximation. It can be observed that a compact subspace of $L = 16$ is sufficient under mild PN conditions, while severe PN requires an expanded dimension to suppress premature saturation, indicating a fundamental trade-off between accuracy and complexity. Furthermore, Fig.~\ref{fig5} demonstrates that the estimated trajectory closely tracks the actual random evolution with high fidelity. This is attributed to the subspace projection eliminating broadband noise while preserving the correlated Wiener structure, enabling precise phase reconstruction even under rapid PN fluctuations.  
	\begin{figure}[tbp]
		\centering
		\includegraphics[width=0.85\linewidth]{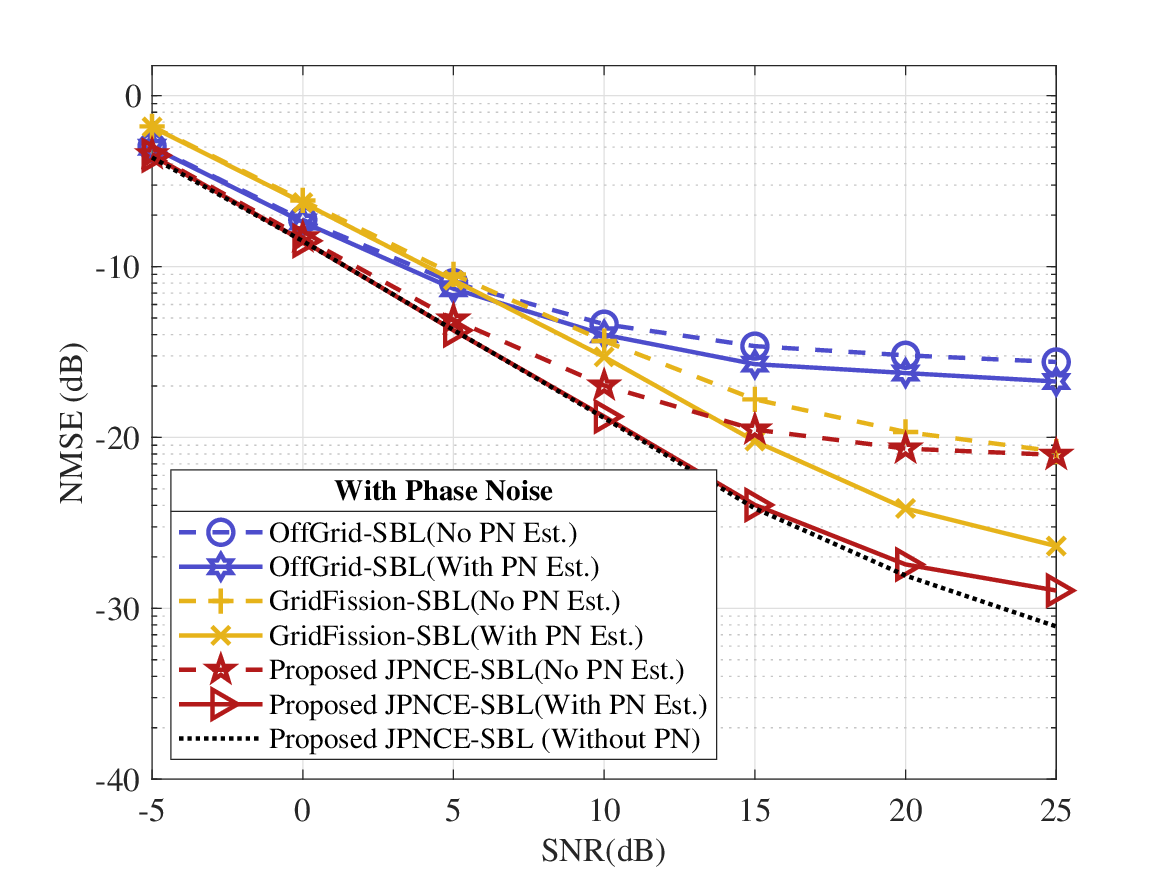}
		\caption{\small{NMSE performance comparison of different estimators versus SNR in the presence of transceiver phase noise.}}
		\label{fig6}
		\vspace{-1em}
	\end{figure} 
	
	From Fig.~\ref{fig6}, it can be observed that when PN is ignored, all estimators suffer from severe error floors at low SNR (around $5$ dB), as the PN-induced phase rotations corrupt the delay-Doppler representation and cause a model mismatch that dominates the thermal noise. When basic PN estimation is introduced, baseline methods such as OffGrid-SBL and GridFission-SBL show only marginal improvement, as they fail to decouple PN variations from the multipath components, leading to early saturation. In contrast, the proposed JPNCE-SBL effectively overcomes this limitation by jointly updating the virtual grid and the PN trajectory, achieving a continuously decreasing NMSE that closely approaches the ideal bound across the entire SNR range.
	\begin{figure}[tbp]
		\centering
		\includegraphics[width=0.85\linewidth]{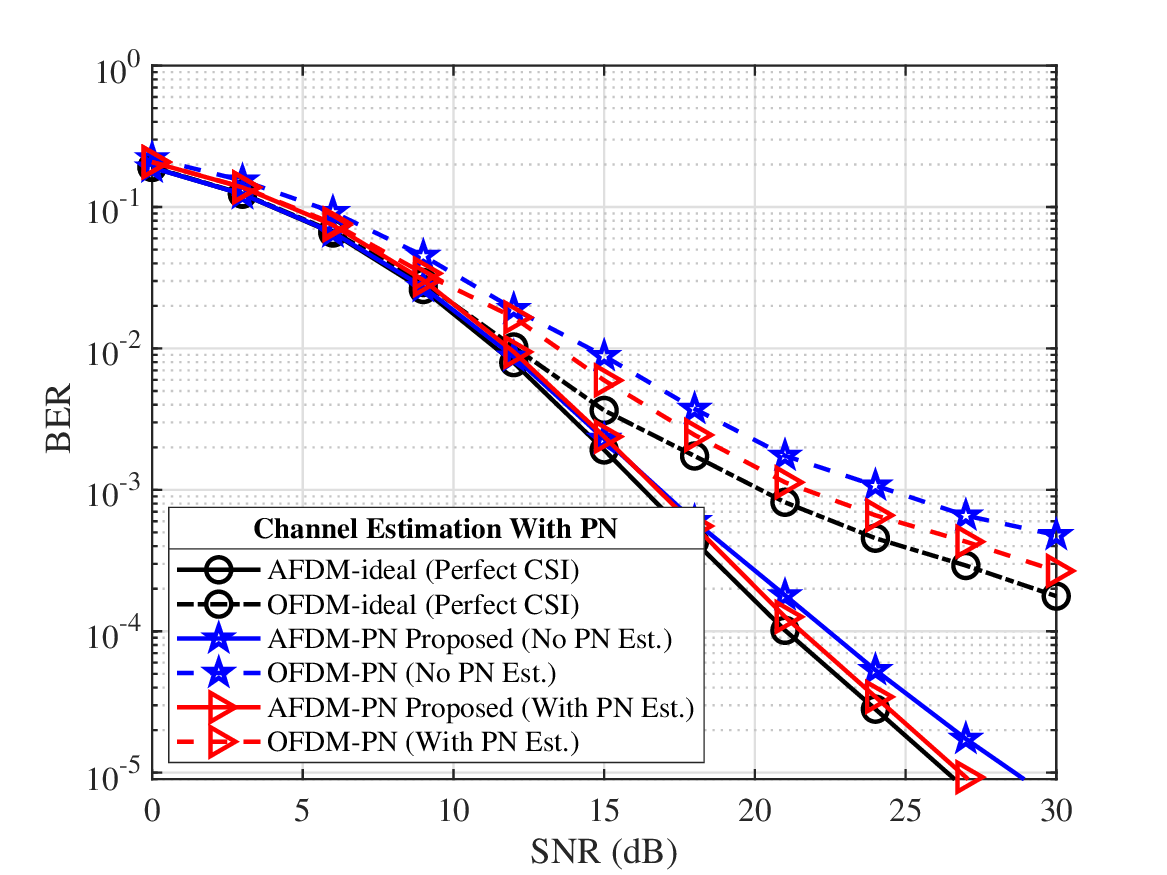}
		\caption{\small{BER performance comparison of different estimators versus SNR in the presence of transceiver phase noise.}}
		\label{fig7}
		\vspace{-1.5em}
	\end{figure}  
	
	From Fig.~\ref{fig7}, it can be observed that under perfect CSI, AFDM outperforms OFDM by fully exploiting the delay-Doppler diversity in doubly-dispersive channels. However, when PN is uncompensated, both waveforms suffer from severe BER saturation, erasing the performance gain of AFDM. In contrast, the proposed JPNCE-SBL recovers this gain by providing high-precision CSI, closely approaching the perfect-CSI bound. This confirms that JPNCE-SBL not only effectively compensates for PN impairments but also maintains the superior performance of AFDM in high-mobility scenarios.
	\vspace{-0.5em}
	\section{Conclusion}\label{sec:conclusion} 
	In this letter, we proposed the JPNCE-SBL framework for joint channel estimation and PN tracking in AFDM systems over doubly-dispersive channels. By incorporating a dynamic grid evolution strategy and a reduced-rank subspace projection mechanism, the proposed framework iteratively refines the continuous delay-Doppler parameters while tracking the time-varying PN trajectory with low computational overhead. Within a unified EM-based probabilistic model, the sparse channel hyperparameters, the PN subspace coefficients, and the virtual Doppler grid are simultaneously updated, thereby effectively preventing error propagation between the two estimation processes. Simulation results demonstrate that the proposed JPNCE-SBL achieves superior NMSE and BER performance compared to existing benchmarks, closely approaching the perfect-CSI bound under practical PN impairments.

\end{document}